\documentclass{iopjournal}

\usepackage[backend=bibtex8,style=numeric,citestyle=numeric-comp,natbib=true,sorting=none,url=false,isbn=false,date=year]{biblatex}
\renewbibmacro{in:}{}

\usepackage{comment}
\usepackage{array}
\usepackage{booktabs}
\usepackage{siunitx}
\usepackage[table]{xcolor}
\usepackage{placeins}

\begin{document}

\articletype{Topical Review}

\title{Constructive community race: full-density spiking neural network model drives neuromorphic computing}

\author{Johanna Senk$^{1,2,*}$\orcid{0000-0002-6304-062X},

Anno C. Kurth$^2$\orcid{0000-0002-9557-1003},

Steve Furber$^3$\orcid{0000-0002-6524-3367},

Tobias Gemmeke$^4$\orcid{0000-0003-1583-3411},

Bruno Golosio$^{5,6}$\orcid{0000-0001-5144-6932},

Arne Heittmann$^7$\orcid{0000-0001-9466-0568},

James C. Knight$^1$\orcid{0000-0003-0577-0074},

Eric M{\"u}ller$^8$\orcid{0000-0001-5880-2012},

Tobias Noll$^9$\orcid{0000-0003-0613-7860},

Thomas Nowotny$^1$\orcid{0000-0002-4451-915X},

Gorka Peraza Coppola$^{2,9}$\orcid{0009-0008-0926-3305},

Luca Peres$^3$\orcid{0000-0001-9748-9073},

Oliver Rhodes$^3$\orcid{0000-0003-1728-2828},

Andrew Rowley$^3$\orcid{0000-0002-2646-8520},

Johannes Schemmel$^{10}$\orcid{0000-0003-1440-4375},

Tim Stadtmann$^4$\orcid{0009-0007-7452-8245},

Tom Tetzlaff$^2$\orcid{0000-0001-5794-5425},

Gianmarco Tiddia$^6$\orcid{0000-0001-7524-0285},

Sacha J. van Albada$^{2,11}$\orcid{0000-0003-0682-4855},

Jos{\'e} Villamar$^{2,9}$\orcid{0009-0007-8791-7100},

Markus Diesmann$^{2,12,13,14}$\orcid{0000-0002-2308-5727}
}

\affil{$^1$Sussex AI, School of Engineering and Informatics,
University of Sussex,
Brighton,
United Kingdom}

\affil{$^2$Institute for Advanced Simulation (IAS-6),
Jülich Research Centre,
Jülich,
Germany}

\affil{$^3$Department of Computer Science,
University of Manchester,
Manchester,
United Kingdom}

\affil{$^4$Lehrstuhl für Integrierte Digitale Systeme und Schaltungsentwurf (IDS),
RWTH Aachen University,
Aachen,
Germany}

\affil{$^5$Department of Physics,
University of Cagliari,
Monserrato,
Italy}

\affil{$^6$Istituto Nazionale di Fisica Nucleare (INFN), Sezione di Cagliari,
Monserrato,
Italy}

\affil{$^7$Neuromorphic Software Ecosystems (PGI-15),
Jülich Research Centre,
Jülich,
Germany}

\affil{$^8$Kirchhoff Institute for Physics,
Heidelberg University,
Heidelberg,
Germany}

\affil{$^9$RWTH Aachen University,
Aachen,
Germany}

\affil{$^{10}$Institute of Computer Engineering,
Heidelberg University,
Heidelberg,
Germany}

\affil{$^{11}$Institute of Zoology, University of Cologne,
Cologne,
Germany}

\affil{$^{12}$JARA-Institute Brain Structure-Function Relationships (INM-10),
Jülich Research Centre,
Jülich,
Germany}

\affil{$^{13}$Department of Psychiatry, Psychotherapy and Psychosomatics, School of Medicine,
RWTH Aachen University,
Aachen,
Germany}

\affil{$^{14}$Department of Physics, Faculty 1,
RWTH Aachen University,
Aachen, 
Germany}

\affil{$^*$Author to whom any correspondence should be addressed.}

\email{j.senk@sussex.ac.uk}


\newpage

\begin{abstract}
The local circuitry of the mammalian brain is a focus of the search for generic computational principles because it is largely conserved across species and modalities. In 2014 a model was proposed representing all neurons and synapses of the stereotypical cortical microcircuit below $1\,\text{mm}^2$ of brain surface. The model reproduces fundamental features of brain activity but its impact remained limited because of its computational demands. For theory and simulation, however, the model was a breakthrough because it removes uncertainties of downscaling, and larger models are less densely connected. This sparked a race in the neuromorphic computing community and the model became a de facto standard benchmark. Within a few years real-time performance was reached and surpassed at significantly reduced energy consumption. We review how the computational challenge was tackled by different simulation technologies and derive guidelines for the next generation of benchmarks and other domains of science.
\end{abstract}

\section{Introduction}

\begin{figure}[h]
    \centering
    \makebox[\textwidth][c]{\includegraphics[width=\textwidth]{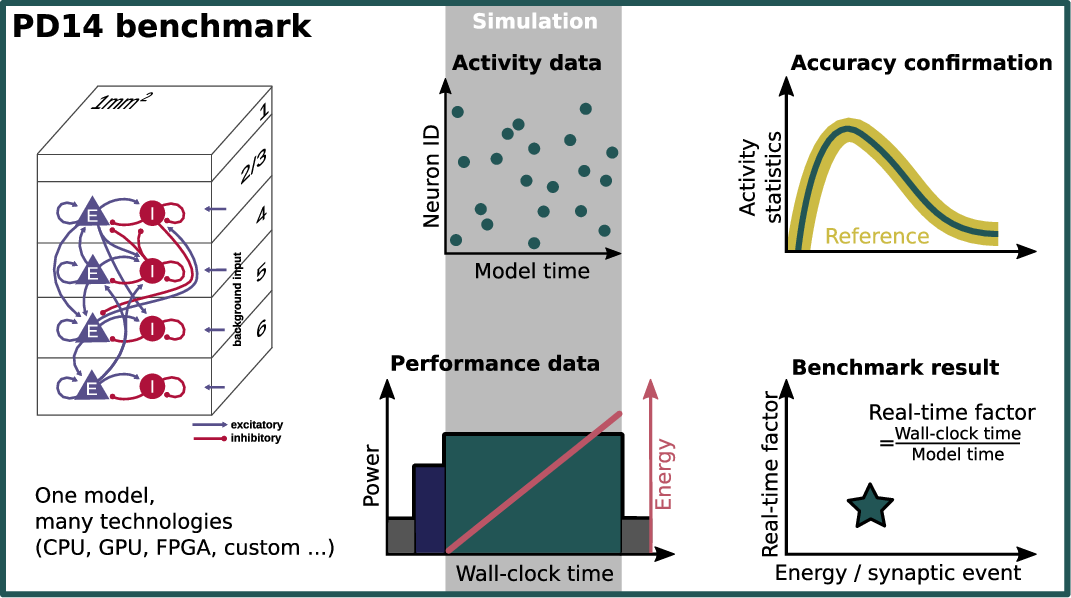}}
    \caption{
    \textbf{Spiking network model of full circuitry below $1\,\text{mm}^2$ surface area of cerebral cortex serves as benchmark for neuromorphic technologies.}
    The statistics (dark cyan curve) of the simulated activity (dark cyan dots) is compared to reference data (thick yellow curve). Once sufficient accuracy is confirmed, the power measured during the simulation phase (light gray background) yields the consumed energy (dark cyan area corresponds to end point of light red curve, dark blue area indicates network construction phase, dark gray idle phase). The performance benchmark result (dark cyan star) contrasts the real-time factor (defined as the ratio of required wall-clock time and biological time covered by the model) against the required energy (expressed as energy per synaptic event). Network sketch (left) reproduced from~\cite{VanAlbada18_291}.}
    \label{fig:graphical_abstract}
\end{figure}

Neuroscience increasingly depends on mathematical modeling and simulations to bridge the gap between the properties of single neurons and the function of the entire brain~\cite{Einevoll19_735}. Thereby, neuroscience moves towards digital twins integrating anatomical and physiological data (e.g., using rat~\cite{Markram2015_456, Reimann22.08.11.503144} or mouse~\cite{Billeh20} models) and  enabling virtual experiments that cannot be done in organisms. A particular structure of interest in this process is the so called canonical microcircuit~\cite{Mountcastle78, Douglas89_480, daCosta10_1265} in the neocortex. The term refers to the local cortical network, the evolutionary youngest part in the mammalian brain~\cite{Florio14_2182}. The microcircuit is canonical in the sense that its basic architecture is conserved across species and modalities~\cite{Douglas04}, making it a natural target for the search of generic computational principles in the brain. In 2014 a model was proposed representing all the nerve cells (neurons) and all their contact points (synapses) below the surface area of $1\,\text{mm}^2$ of cortex~\cite{Potjans14_785}; that is a network size of on the order of one hundred thousand neurons and one billion synapses. This removed all uncertainties about the effects of downscaling on network activity in earlier models~\cite{Albada15}. Downscaling here refers to the concept of reducing the number of neurons and synapses in a model for computational tractability, while attempting to preserve the dynamics and keeping other structural characteristics like the probability of two neurons forming a connection invariant. It does not refer to representing neurons and synapses at a finer level of description, like the level of electrical compartments or the molecular level. The model represents four cortical layers by populations of excitatory and inhibitory leaky integrate-and-fire model neurons (network sketch in Fig.~\ref{fig:graphical_abstract}). Due to the simplicity of its components and the limited data available at the time, the explanatory scope of the model is limited, even though it reproduces basic features of cortical activity such as asynchronous and irregular spiking at biologically plausible population-specific rates. 
Initially the simulation results were confirmed by independent researchers using different simulation codes~\cite{Shimoura18_ReScience,Romaro21_1993}. The model and variations thereof have been used for various neuroscientific investigations (e.g., attention~\cite{Wagatsuma11_00031, wagatsuma2013spatial}, orientation selectivity~\cite{Merkt19_e1007080}, inhibitory neuron types~\cite{Lee17_28,wagatsuma2023_4459,Jiang24_378}, network connectivity~\cite{Diaz16_57,Giacopelli21_4335}, building block of multi-area models~\cite{Schmidt18_e1006359,pronold2024clustered}, forward-modeling of extracellular potentials~\cite{Hagen16_bhw237,Naess21_117467}) and as a test bed for theoretical methods (e.g., mean-field and linear response theory~\cite{Bos16_1,Schuecker17}, population density models~\cite{Cain16_e1005045,Osborne21_159}, stochastic population equations~\cite{Schwalger17_e1005507, Rene2020_1448}).
Nevertheless, when it was first proposed, run times of several minutes were required for the simulation of $1\,\text{s}$ of biological time, limiting the practical use of the model. A recent review \citep{Plesser25_295} analyses the impact of the model on the fields of computational neuroscience and neuromorphic computing.

The opportunity to overcome a barrier in terms of network size, and the availability of a reproducible model combined with the computational challenge of bringing down simulation time of a microscopically parallel problem sparked a concrete and quantifiable race in the computational neuroscience and neuromorphic computing communities for ever faster and more energy-efficient simulation. In particular the frequent updates (at least every $0.1\,\text{ms}$ of model-time in time-driven simulations) as well as the communication and delivery of a large number of events (also called spikes) during state propagation call for innovative solutions in hardware and software, beyond conventional approaches in High-Performance Computing (HPC).
A few years later, the model has been simulated on a range of platforms including a large-scale neuromorphic system (SpiNNaker~\cite{VanAlbada18_291,Rhodes19_20190160}), many-core CPU systems (NEST~\cite{VanAlbada18_291,Kurth_2022}) GPUs (GeNN~\cite{Knight18_941,Knight21_15}, NEST GPU~\cite{Golosio21_627620,Golosio23_9598} and FPGAs (CsNN~\cite{heittmann22_728460}, neuroAIx~\cite{kauth23_1144143}). In this time span, real-time performance was reached and simulation time and energy per synaptic event dropped by two orders of magnitude. The circuit can now be simulated an order of magnitude faster than real time.
The role of the model as a benchmark was never declared by any organization but emerged over the years. The studies reproducing the work were done independently by different laboratories. It is therefore time to gather the results distributed in the literature, contrast the findings and identify the advantages and bottlenecks of different approaches. From this we derive limitations of present technology and requirements for future neuromorphic architectures.
From the perspective of computational science we discuss what made the PD14 model successful in becoming a de facto standard benchmark for our community and what obstacles the authors faced using the benchmark. 
Finally, we discuss next challenges for the simulation of large-scale neuronal networks and characterize desired properties of next generation benchmark models.

\section{Results}
\label{sec:results}

This work covers technologies designed for large-scale spiking neuronal network simulations. These technologies have different algorithmic approaches and include software-based simulators exploiting conventional hardware and custom neuromorphic platforms. Their performance is assessed on the example of the PD14 model~\cite{Potjans14_785} in terms of time to solution and energy to solution. At this point, we take it for granted that the model is simulated at sufficient accuracy, meaning that the statistics of the simulated spike data are compatible with reference data (Fig.~\ref{fig:graphical_abstract}, upper panels, for details see Section~\nameref{sec:recipe_accuracy}). Timestamps in the simulation scripts enable different execution phases to be identified and related directly to temporally resolved power measurements. (Fig.~\ref{fig:graphical_abstract} lower panels). Each simulation experiment begins with a network initialization phase during which the network is constructed either directly on the simulation system or on a host system and then transferred. The beginning of the subsequent state-propagation phase should be considered a warm-up time because start-up transients may occur due to initial conditions. Even though network initialization and warm-up time can consume a substantial amount of time and energy, here we only focus on the state-propagation phase with stationary network dynamics. We define the performance metrics as follows:
\newpage

\begin{paragraph}{Time to solution}
The \textit{real-time factor} $q_\text{RTF}$ is defined as the quotient of wall-clock time $T_\text{wall}$ (which is also known as real time) and model time $T_\text{model}$ (which is the duration by which the state of the model is advanced in time):
\begin{equation}\label{eqn:rtf}
    q_\text{RTF} = \frac{T_\text{wall}}{T_\text{model}.}
\end{equation}
If the real-time factor is larger than one, the simulation runs slower than wall-clock time. The reciprocal of $q_\text{RTF}$ is a measure for the speed of the calculation; the higher $q^{-1}_\text{RTF}$ the faster the computing system completes the task. Ten times faster than real time could be expressed by $q^{-1}_\text{RTF}=10$.

The speed of the calculation should not be confused with the concept of speedup. Speedup~\cite{Wilkinson04,vanAlbada21_47} refers to a ratio of two wall-clock times, namely the wall-clock time required to solve a problem serially and the wall-clock time needed to solve the same problem in parallel (e.g., by distributing the work across multiple threads or processes): $q_\text{speedup} = T_\text{wall,serial} / T_\text{wall,parallel}$.
\end{paragraph}
     
\begin{paragraph}{Energy to solution}
The \textit{energy per synaptic event} $E_\mathrm{syn}$ is defined as the integrated power $P$ (i.e., wall socket power including communication infrastructure) over the state-propagation phase (i.e., wall-clock time $T_\text{wall}$) divided by the total number of synaptic activations in the time interval $T_\text{model}$:
\begin{equation}\label{eqn:num_spikes}
    E_\text{syn} = \frac{\int_0^{T_\text{wall}}P(t)\,\text{d}t}{T_\text{model} \cdot \sum_\alpha N_\alpha \cdot K_{\text{out},\alpha} \cdot \nu_\alpha}
\end{equation}
with the neuron number $N_{\alpha}$, the average number of outgoing connections $K_{\text{out},\alpha}$ and the spike rate $\nu_\alpha$ per neuron of population $\alpha \in \{ \text{L2/3E, L2/3I, L4E, L4I, L5E, L5I, L6E, L6I} \}$ in the PD14 model. In a simulation of ten seconds of model time our neuronal network produces approximately 2.47 million spikes leading to approximately 9.6 billion synaptic events.
\end{paragraph}
\bigskip

The PD14 has become a de facto standard benchmark. In the beginning, an executable model description was only available in the original simulation language (SLI) of the NEST code~\cite{Gewaltig_07_11204} and this has been included in all versions of NEST since v2.4.0 (released in 2015). But, thanks to progress in the community and large-scale projects, executable descriptions abstracted from a particular simulation engine became available. Nevertheless, a performance result is only meaningful if a benchmark reaches the same accuracy of the solution as the reference data. This is because generally, a lower accuracy can be reached with less computational effort. The neuroscientific purpose of the PD14 model is to reproduce certain characteristics of the activity of the respective biological neuronal network. Therefore, \citet{VanAlbada18_291} define a set of measures a benchmark of PD14 needs to fulfill. A later study~\cite{Dasbach21_90} finds that the specified correlation measure is good enough to expose corrupted activity, but that model-specific correlation structure is only exhibited at considerably longer simulation times. This study also shows that the inhomogeneity of the network structure hides the distribution of synaptic weights: the correlation measures remain identical if all weights are collapsed to the mean value of the corresponding normal distribution. Retrospectively this is apparent as a mean-field theory~\cite{Bos16_1} can quantitatively reproduce the first and second-order statistics of the population activity. Thus, the characterization of the solution does not require the simulation of a spiking neuronal network with an individual representation of all neurons and synapses. The synaptic weights can be replaced by a single value per neuron population~\cite{Dasbach21_90} and connections can be generated on-the-fly with a generator for pseudo-random numbers~\cite{heittmann22_728460}. This dramatically reduces the amount of memory a simulation engine needs to access. Nevertheless, so far only direct simulation delivers the full distributions of the statistical measures.
It remains to the good will of the scientists to carry out the benchmark with simulation engines that are in principle capable of representing synaptic weights as dynamical variables as required for plasticity and learning. Spike-based plasticity rules directly interact with the correlation structure. If constraints of the engine at hand do not allow a representation at this level of resolution, the results may still be interesting but the conditions need to be declared.

The community gain of the PD14 race is attributed to various factors and a complete disentanglement is outside the scope of this manuscript. General-purpose approaches (i.e., simulation software designed to support a plethora of different models and to run on different conventional CPU and GPU based hardware systems such as NEST~\cite{Gewaltig_07_11204} and GeNN~\cite{Yavuz16_18854}) are compared to dedicated approaches (i.e, neuromorphic hardware such as SpiNNaker~\cite{furberpetrut} or software-hardware-model co-designs using FPGAs). Apart from model-specific simulator advancements, some progress in case of NEST and GeNN is due to later software versions and hardware generations which are continuously optimized independently of the PD14 model. However, the proportion of the contributions is unclear as both are replaced in new data points. For SpiNNaker instead, we see an improvement of the software stack only allowing better use of the existing hardware. Generally speaking, the observed difficulty of running old software on new hardware hampers a systematic comparison. For example, despite great efforts to adhere to programming language and coding standards, in practice it is a challenge to run NEST versions released only a few years ago on recent hardware. Further research may analyze why this is so difficult and what developers of a simulation software stack and developers of benchmark models can learn to make simulation codes and models more robust.

\begin{table}[h]
\centering
\begingroup
\definecolor{darkcyan}{HTML}{225555}
{\rowcolors{1}{darkcyan!10}{darkcyan!20}
\renewcommand{\arraystretch}{2}
{\scriptsize
\begin{tabular}{p{1.1cm}p{0.5cm}|S[table-format=3.2]|p{0.9cm}|p{1.6cm}|>{\raggedleft\arraybackslash}p{1cm}|p{3.2cm}|>{\raggedleft\arraybackslash}p{0.8cm}|p{1cm}}
\toprule
Study &  & $q_\mathrm{RTF}$ & $E_\mathrm{syn}$ ($\mu$J) & Simulator & \#Nodes & System & Process node (nm) & External drive \\
\midrule
vAl+18a & \cite{VanAlbada18_291} & 2.465 & 9.941 & NEST CPU & 12 & 2 Intel Xeon E52680v3 & 22 & DC \\
vAl+18b & \cite{VanAlbada18_291} & 4.584 & 5.816 & NEST CPU & 3 & 2 Intel Xeon E52680v3 & 22 & DC \\
vAl+18c & \cite{VanAlbada18_291} & 20 & 4.4 & SpiNNaker 1 & 6 & 48 x 18 x ARM-968 & 130 & DC \\
KN18 & \cite{Knight18_941} & 1.838 & 0.47 & GeNN & 1 & Tesla V100 & 12 & Poisson \\
Rho+19a & \cite{Rhodes19_20190160} & 1 & 0.601 & SpiNNaker 1 & 12 & 48 x 18 x ARM-968 & 130 & DC \\
Rho+19b & \cite{Rhodes19_20190160} & 1 & 0.628 & SpiNNaker 1 & 12 & 48 x 18 x ARM-968 & 130 & Poisson \\
Gol+21 & \cite{Golosio21_627620} & 1.055 & 0.25 & NEST GPU & 1 & RTX 2080 Ti & 12 & Poisson \\
Kni+21 & \cite{Knight21_15} & 0.7 & -- & GeNN & 1 & Titan RTX & 12 & Poisson \\
Hei+22 & \cite{heittmann22_728460} & 0.25 & 0.284 & CsNN & 345 & IBM INC-3000 & 28 & Poisson \\
Kur+22a & \cite{Kurth_2022} & 0.53 & 0.48 & NEST CPU & 2 & 2 AMD EPYC Rome 7702 & 7 & DC \\
Kur+22b & \cite{Kurth_2022} & 0.67 & 0.33 & NEST CPU & 1 & 2 AMD EPYC Rome 7702 & 7 & DC \\
Gol+23a & \cite{Golosio23_9598} & 0.386 & 0.104 & NEST GPU & 1 & RTX 4090 & 5 & DC \\
Gol+23b & \cite{Golosio23_9598} & 0.272 & 0.074 & GeNN & 1 & RTX 4090 & 5 & Poisson \\
Kau+23 & \cite{kauth23_1144143} & 0.05 & 0.048 & neuroAIx & 35 & NetFPGA SUME & 28 & DC \\
\bottomrule
\end{tabular}
}}
\endgroup
\caption{Performance data of considered studies. A code name disambiguates the studies and refers to the bibliography (number in brackets). The real-time factor $q_\text{RTF}$ (Eq.~\ref{eqn:rtf}) and the energy per synaptic event $E_\text{syn}$ (Eq.~\ref{eqn:num_spikes}) are the performance results obtained with the simulation technologies characterized by simulator name, number of nodes, system specification, and the process node as the industrial specification of the chip technology. The external drive indicates the background input used (DC or Poisson).}
\label{tab:performance}
\end{table}

\begin{figure}[h]
    \includegraphics[width=\textwidth]{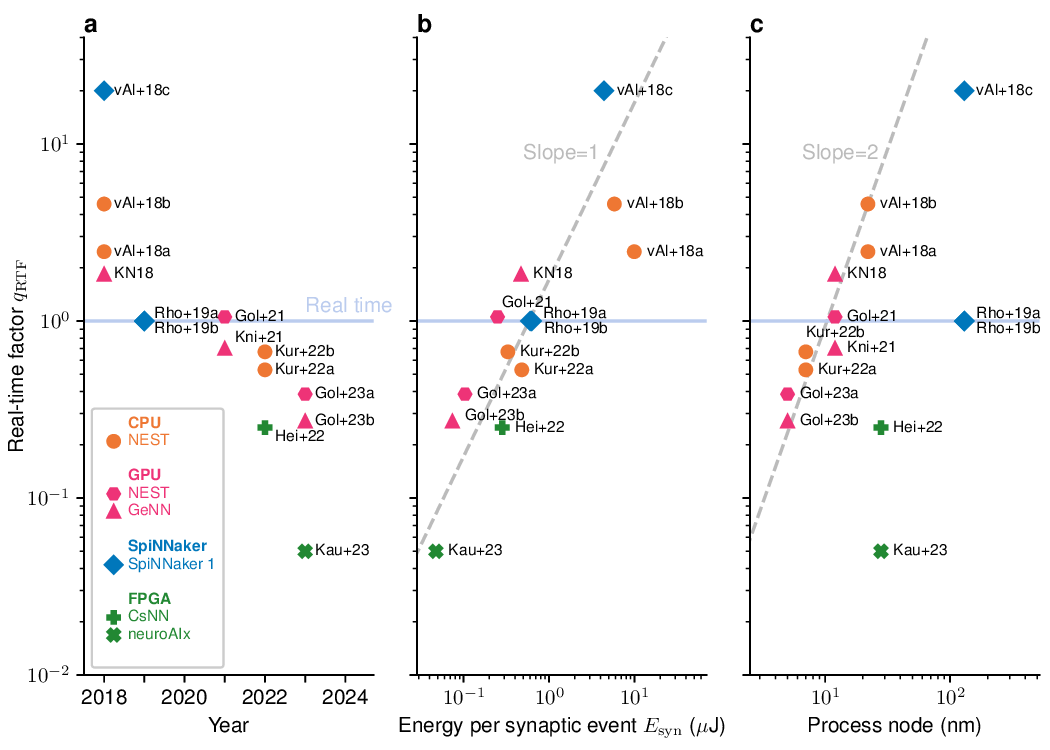}
    \caption{
    \textbf{Progress of the community in reduction of time to solution and energy consumption for the PD14 model.}
    Colors group hardware architectures and shapes indicate algorithmic approach (legend). Abbreviations in panels further disambiguate individual studies.
    \textbf{a}~Ratio between time passed on wall-clock and stretch of time covered by the model (real-time factor) versus the year of publication in semi logarithmic representation.
    \textbf{b}~Real-time factor as a function of energy per synaptic event in double logarithmic representation. Dashed line from fit through all data points with a slope of one.
    \textbf{c}~Real-time factor versus process node in double logarithmic representation. Dashed line from fit through CPU and GPU data points with a slope of two.
    Citations of studies and values are given in Table~\ref{tab:performance}.
    }
    \label{fig:performance}
\end{figure}

\begin{figure}[h]
    \centering
    \includegraphics[width=\textwidth]{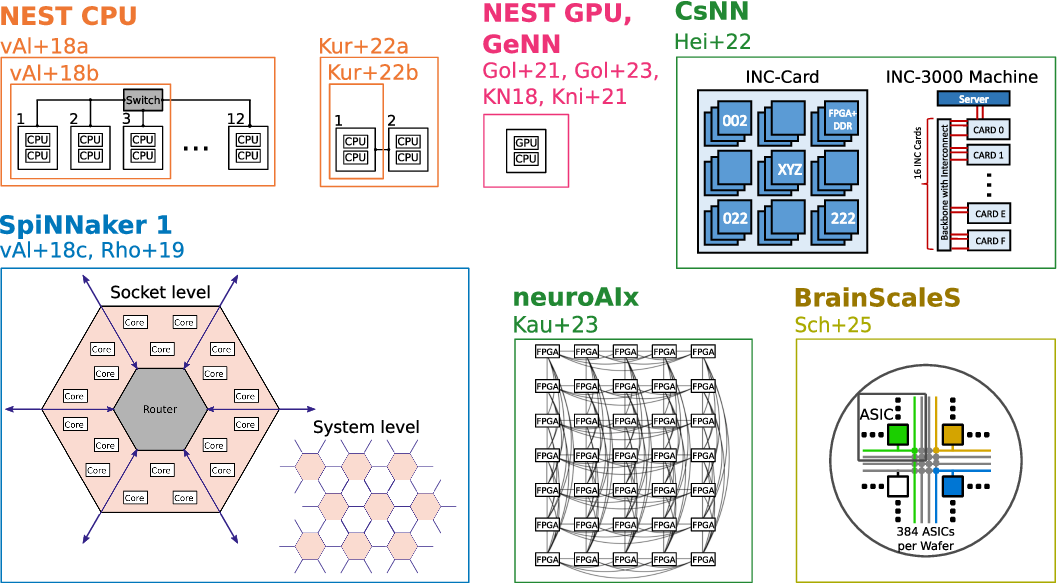}
    \caption{
    \textbf{Sketches of computing platforms.} Diagrams illustrate the major computational building blocks and the communication architecture of a particular system required during the simulation phase.
    Compare with columns ``\# Nodes" and ``System" in Table~\ref{tab:performance}. Same color scheme as in Fig.~\ref{fig:performance}.
    \textbf{NEST CPU} uses compute nodes with two CPU sockets each. The first study (\emph{vAl+18a/b}) uses a compute cluster with Infiniband switch. The second study uses two point-to-point connected nodes (\emph{Kur+22a/b}).
    \textbf{NEST GPU} and \textbf{GeNN} simulations run on single GPUs.
    \textbf{SpiNNaker} uses a router and 18 low-power cores per socket, with sockets embedded in a mesh routing fabric with 6 links per socket, enabling an extensible system facilitating multicast communication between cores. The presented studies using SpiNNaker both used this architecture, with variations in how the cores were configured to perform neural processing, and small variations in the overall size of the systems.
    \textbf{CsNN} uses the IBM INC-3000 system which consists of 16 PCB's (INC-cards), each hosting 27 reconfigurable SoC nodes. The whole INC-3000 system consists of 432 FPGA nodes which are connected by a 3-dimensional mesh.
    \textbf{neuroAIx} uses a cluster of 35 FPGAs connected with overlayed long hop topologies as an extension to mesh-like network topologies.
    \textbf{BrainScaleS} uses a circuit-switched on-wafer network for spike communication between the 384 ASICs per wafer.
    }
    \label{fig:systems}
\end{figure}
\FloatBarrier

Table~\ref{tab:performance} and Fig.~\ref{fig:performance} summarize the progress of the community in increasing the simulation speed and reducing the energy consumption of the PD14 model. The employed computing platforms are illustrated in Fig.~\ref{fig:systems}. The historical account (Fig.~\ref{fig:performance}a) of the real-time factor starts with a number of initial data points in 2018. None of the studies came close to real time. The SpiNNaker system is specified for real-time performance, but shortcomings of the mapping of neurons to hardware exposed by the study required a temporary slow down. The CPU code, already mature at the time, showed better performance but at the highest energy consumption. GPU code for spiking neuronal networks was just emerging and immediately took the top position in speed. A year later the SpiNNaker system reached its specification and simulated the model in real time. In the following three years progress was harder and breaking the real-time barrier became a psychologically important target. Finally a new GPU code came very close and the more mature initial GPU code passed the barrier. Shortly afterwards CPU code outperformed the GPU results and reached out into the real-time regime on a single state-of-the-art commodity compute node. Nevertheless this was dwarfed in the same year by an FPGA-based neuromorphic system. A year later the rapid progress in GPU hardware enabled GPU-based systems to take the lead again. But in the same year a new dedicated FPGA-based neuromorphic system made a jump to a simulation speed twenty times faster than real time. This achievement is substantial because present CPU code requires a longer duration just for the communication between compute nodes.

In the real-time factor alone (Fig.~\ref{fig:performance}a) the progress appears steep but continuous with the exception of a major jump due to the FPGA technology. The vertical divide at real time may indicate the psychologically important barrier motivating further publications. Fig.~\ref{fig:performance}b shows the time required to reach the solution as a function of the energy required by the different systems to complete the task. The double-logarithmic representation reveals an overall power-law decline. While energy consumption dropped by two orders of magnitude, the simulations became two orders of magnitude faster. The graph does not segregate technologies but data points are rather organized by the time of publication.

Fig.~\ref{fig:performance}c shows the real-time factor over process node. Even though the communication hardware and other system-specific components are not resolved, this representation reveals a clearer trend and at the same time a separation of dedicated hardware architectures from the conventional CPU and GPU architectures. Let us assume that while reducing process node, manufacturers increase the number of transistors and keep the chip area constant, thereby also conserving the power consumption; keeping the power density constant is known as Dennard scaling~\cite{Dennard74_256}, though such scaling has been increasingly inapplicable to digital processing systems beyond the $65\,\text{nm}$ technology node~\cite{Bohr07_11}. If neuromorphic code keeps up with the increased parallelization~\cite{Aimone23_418} enabled by the growing number of transistors, the real-time factor should drop linearly with the number of transistors. The latter, however depends quadratically on process node. Following this argument we expect in the double-logarithmic representation of Fig.~\ref{fig:performance}c the real-time factor to decline with a slope of two in dependence of process node. This holds well for the conventional CPU and GPU technologies even though the impact of the whole system used is disregarded in this argument. The argument also explains the slope of one in the dependence of real-time factor on energy per synaptic event observed in Fig.~\ref{fig:performance}b. If a technology completes the task in half the time, only half of the energy is required under the assumption of a constant power density. The dedicated architectures based on the SpiNNaker chip and FPGAs deviate from this trend. The expectation on a dedicated hardware is that for comparable process node it outperforms code executed on a conventional computing architecture. Initially the SpiNNaker system (\emph{vAl+18c}) suffered from a sub-optimal mapping of PD14 onto the hardware. But already at this time performance was better than the one predicted for CPU and GPU based systems at this large process node. Conversely, starting from the \emph{vAl+18c} data point and projecting performance with a slope of two down to the process node of the conventional competitors at the time (\emph{vAl+18a/b}) also exposes the superior performance of the design. With improved software a dramatic step down in real-time factor became possible (\emph{Rho+19}) while process node remained unchanged. The architecture is competitive with conventional architectures using an order of magnitude smaller process node. Furthermore, a prediction starting at (\emph{Rho+19}) with a slope of two down to smaller process nodes outperforms all recent CPU and GPU systems. The same is true for the FPGA-based systems. They already started out in the range of the best real-time factors (\emph{Hei+22}). A system of generic FPGA boards with customized communication and selective optimization in Hardware Description Language (HDL) yielded a step down in real-time  factor by an order of magnitude while the process node remained the same (\emph{Kau+23}). With the optimized software the SpiNNaker and FPGA based systems fall on the same projection for decreasing process node. 
Following the trend of the conventional architectures for further reductions in process node it seems difficult to reach $q_\text{RTF}=0.1$ without major progress in algorithms. Applying the same projection with a slope of two to the SpiNNaker and FPGA architectures, however, $q_\text{RTF}=0.01$ seems indeed plausible.

The reduction by two orders of magnitude in simulation time since the beginning of the race enables the community to investigate plasticity and system-level learning in a full scale model where neurons are supplied with the natural number of synapses. This increase in simulation speed is of practical relevance for the number of simulation experiments a researcher can do per day, because the biological processes unfold over minutes and hours. Trivial parallelization is often not a cure here, because only the result of a simulation gives the insight and inspiration for the next one. On a more societal level, the energy costs and respective CO$_2$ footprint of simulating the PD14 model are reduced.

Unless a neuromorphic system needs to interact with the real world in a closed loop~\cite{Mahowald91_515}, it is worth to strive for even faster and more energy efficient simulations. For the use as a neuroscience research tool, which requires a high degree of flexibility, $100\times$ faster than real-time is desirable and seems to be compatible with present day conventional semiconductor technology. This was also concluded in a previous pilot project (Advanced Computing Architectures (ACA), \url{https://www.fz-juelich.de/en/aca}; e.g., \cite{Kauth20_1,Trensch22_33}). For the PD14 this may constitute the next milestone for the community. 

For conventional CPU and GPU technology, research is ongoing on improving the codes for multi-node systems. 
But for $100\times$ the time available for an update step of $0.1\,\text{ms}$ is already shorter than the latency of an older Infiniband interconnect ($5\,\mu\text{s}$), and close to the latency of a modern one ($0.5\,\mu\text{s}$). The progress on more densely integrated single nodes, however, is steady and the success of strong-scaling on multi-GPU nodes is largely unexplored. In this respect, the booster nodes of the upcoming JUPITER~\cite{Herten24_1} exascale computer at Juelich equipped with four NVIDIA GH200 Grace Hopper Superchips are an exciting next target for CPU as well as GPU code. However, GPU codes typically need to launch multiple kernels during each simulation timestep to ensure correct synchronisation and each launch incurs a fixed latency (a mimimum of $2.5\,\mu\text{s}$ was measured on an NVIDIA DGX-1 system~\cite{Zhang19_5}). Therefore, as for example GeNN launches four kernels per timestep when simulating PD14, this launch latency alone limits $q_\text{RTF}$ to around $0.1$.

SpiNNaker~1 uses a relatively old semiconductor technology, 130\,nm bulk CMOS, which should be taken into consideration when reviewing energy performance, but further highlights the importance of overall system architecture on optimising execution speed of spiking neural network simulations. The successor chip, SpiNNaker~2~\cite{mayr2019spinnaker}, has been developed using a more up-to-date technology (22\,nm FDSOI) and is expected to deliver around a $10\times$ improvement in functional density and energy-efficiency. As established above, extrapolating the FPGA architecture with a slope of two  shows that $100\times$ is in reach of reasonable process node sizes. Critical is of course a corresponding communication architecture~\cite{Kauth20_1}.

There is a convergence of general purpose computational architectures and dedicated neuromorphic hardware in the sense that in conventional systems the parallelism becomes ever more fine grained. For the PD14 model for example, each of the $128$ cores of the compute nodes used here~\cite{Kurth_2022} is just responsible for some 700 neurons. It remains to be seen how far this abundance of hardware can be exploited for further strong scaling. The structures in present day chips are orders of magnitude smaller than the respective structures in biology and the switching times are accordingly shorter. Therefore electronics can profit from techniques like busses and multiplexing to compensate for the lack of cables. Nevertheless, the optimal system is not necessarily one where one circuit takes care of a single neuron. The BrainScaleS architecture (Fig. \ref{fig:systems} and section \nameref{sec:brainscales}) exploits the microscopically parallel nature of neuronal networks to the extreme. Each neuron is emulated using its own analogue circuit and achieves $q_\text{RTF}=0.0001$ by mapping the full network to a single wafer-size chip. At present only a downscaled version of PD14 fits onto the system~\cite{Schmidt25_nice}, therefore the data are not included in Fig.~\ref{fig:performance}. Progress in the combination of analog and digital hardware as well as mapping software may change this.

Projecting the linear dependence of the FPGA architecture in Fig.~\ref{fig:performance}b to $q_\text{RTF}=0.01$ would lead to an energy consumption per synaptic event of approximately $10\,\text{nJ}$.
BrainScaleS already reaches the same order of magnitude ($12\,\text{nJ}$) for the downscaled version of PD14.
This is three orders of magnitude lower than the value found in the initial studies in 2018, but still five to six orders away from the amount required by nature ($19$ - $760\,\text{fJ}$, see~\cite{VanAlbada18_291}).

Since the PD14 model was initially developed with a focus on neuroscience, there were uncertainties about model and simulation parameters to be used for benchmarks. For example, whether to use DC or Poisson input as external drive may affect the performance and is therefore indicated in Table~\ref{tab:performance}. We identify the choice of initial conditions, treatment of a warm-up time, simulation duration, repeated simulations, and spike recording as further settings that are inconsistently handled across the studies involved and influence comparability (for details see \nameref{sec:studies}). In fact, the effect of certain model and simulation parameters on network activity and compute requirements was only exposed one at a time as byproduct of the unstructured community effort. To pool the gained insights and provide the missing instructions we here compile a \nameref{sec:recipe} with recommendations for future benchmark simulations including a model reference with detailed simulator-independent model documentation.

Most of the studies used different, hand-crafted model implementations suited to their simulators. Developing and testing these implementations is a laborious and error-prone task. With the benchmarking recipe proposed here, we aim to facilitate obtaining and porting the reference implementation and conducting the benchmarking experiments. Additionally, they highlight a number of snares impeding the communication between researchers and increasing the difficulty of comparing the results. To further simplify developing and using such benchmarks, a common language similar to the simulator-independent language PyNN~\cite{Davison09_10} with backends for different systems as well as shared definitions for the measured quantities and a unified verification and validation workflow with distinguished reference data could be instrumental for the community. Another challenge lies in unifying benchmarking tools for running simulations and comparing results, addressing inherent but often overlooked issues in performance benchmarking~\cite{Albers22_837549}.

\section{Discussion}
\label{sec:discussion}

The race is ongoing but the goals to achieve a simulation speed one hundred times faster than real time and to rationalize the comparison of different architectures comes into sight. As the number of computational elements converges towards the number of neurons it seems as if the von-Neumann bottleneck widens even for conventional computing systems.

Finding challenging benchmarks remains relevant for the validation of dedicated neuromorphic computing systems. The construction of such a system is justified if it outperforms in some measure the conventional mass produced CPU and GPU based systems equipped with the best possible software. Fundamental limitations showing the superiority of neuromorphic approaches in strong-scaling or weak-scaling scenarios would give confidence and guidance for technology development.

The unfolding race highlights conditions enabling and motivating such a community wide benchmarking effort but also exposes obstacles in interpreting the benchmark definitions and initial weakness in the metrics for the verification of the simulation results.

Unlike other benchmarks in the area of brain-inspired computing, our metrics for comparing the performance of different approaches are not addressing the accuracy achieved in solving a particular task. It is already part of the verification of a certain system and a precondition that a minimal accuracy is met. The actual comparison is on the time it takes to reach the solution (at required accuracy) and the energy consumed. 

The proposed \nameref{sec:recipe} is a step towards more consistent benchmark runs. 
Porting the model to a new hardware or software environment, however, remains a challenging task due to the inherent complexity of benchmarks~\cite{Albers22_837549} that requires in-depth understanding of the model structure, dynamics, and the respective computing platform. Together with the model reference we publish the recipe as a living document such that the guidelines can be improved or extended if further uncertainties emerge. We also include the code to generate the performance figure (Fig. \ref{fig:performance}) such that new data can be included.

We believe that the insights on the conditions for the success of the study and the origins of difficulties are not specific to our field computational neuroscience but generic for benchmarking in computational science. The benchmarking recipe condenses this knowledge. In short, it highlights the importance of a well-documented model description with a reference implementation and simulation instructions.

Network initialization can require substantial amounts of resources, especially when exploring various parameters. The authors believe that the software stacks of the studies reviewed here are not yet mature enough for meaningful comparisons.
Therefore our investigation concentrates on the phase of state propagation. The longer the stretch of biological time covered by the model, the more is the complete time to solution dominated by the state propagation phase. Nevertheless, with maturing software stacks future work should also include the time and energy required for network construction.

Since the PD14 model's conception, new data sets have emerged, expanding our neuroscientific understanding of cortical microcircuits. While model revisions could incorporate additional insights (e.g., target-type specific innervation preferences depending on whether a neuron is excitatory or inhibitory ~\cite{Kurth24_biorxiv}, distinct interneuron types and short-term plasticity~\cite{Jiang24_378}, and point-neuron vs. multi-compartment neurons~\cite{Billeh20}), they are not discussed here as they would not significantly alter the computational load.
Still we argue for a transparent and systematic tracking of model revisions as done with the PD14 model within the NEST code over the past ten years and from now on in a self-contained \nameref{sec:recipe_reference} repository.

Besides, the complexity of the brain calls for complementary benchmark models with different biological detail and computational demands such that large-scale neuromorphic hardware systems will not be optimized for only a single model type. The benchmark considered here represents the smallest building block of the part of the mammalian brain responsible for higher brain functions. The strong-scaling scenario has exposed the characteristics of the hardware systems and their limits. It is now time to bring the weak-scaling scenario into perspective. For our research field this means capturing larger parts of the brain at the same level of resolution. One way of achieving this is not to increase the surface area of a model of a single cortical area, but to consider interconnected patches of surfaces in several areas. This is neuroscientifically of interest because it relates the activity in local circuits to the global dynamics of the brain.
The workflows for porting a neuronal network between computing platforms are now established. This should make it easier to carry out and compare benchmarks for other, more advanced, networks models. An example is a multi-area model of the visual system of macaque monkey using an adapted version of the PD14 model for each of the $32$ areas represented~\cite{Schmidt18_1409}. This model can already be simulated with NEST CPU~\cite{Schmidt18_e1006359}, NEST GPU~\cite{Tiddia22_883333}, and GeNN~\cite{Knight2021_136} but speed is far from real time. Another model adaptation incorporating distance-dependent connectivity has not been simulated on other systems than NEST CPU yet~\cite{Senk24_405}.

More detailed neuron models and slow processes like long-term plasticity, system-level learning, and development increase the computational costs of any model. Thus, at the scale of PD14 we will face longer simulation times than predicted by the data shown here, and plastic processes still cannot be studied in larger models even with state-of-the-art simulation engines. Another challenge therefore consists in devising and maturing additional representative benchmark models which require different metrics for evaluating the accuracy. It is here where benchmarks concentrating on time to solution and energy to solution converge with benchmarks concentrating on the performance in fulfilling a function~\cite{Yik25_1}.

Much like the supercomputing community has learned that a useful benchmarking suite needs to be multi-disciplinary for a multi-purpose system (see for example~\cite{Herten24_1}). At the same time not all benchmarks are of relevance for all systems. A chip for edge computing has a different area of application than a chip designed to simulate a cortical network. The field of neuromorphic computing needs to map out its application domain and arrive at a consistent set of benchmarks ~\cite{Yik25_1}. The optimal application area of a given approach can then be characterized as a particular territory on this map.

Computational neuroscientists and simulation system developers symbiotically benefit from diversity in both the types of neuronal network models studied and the simulation technologies developed. On the road towards understanding the brain and at the same time making use of the gathered knowledge to advance technology, we realized the importance of points of convergence for different disciplines to come together and learn from each other through rigorously defined and executed performance benchmarks. Accepting the effort and extra complexity of using a real-world benchmark instead of a synthetic benchmark, sometimes also called mini-app, has the advantages of reducing the danger that important characteristics are overlooked and that the relevance of the benchmark is not in question.
We hope that the experiences shared here advances computational science beyond the neuroscience domain.

\section{Methods}
\label{sec:methods}

\subsection{Technology and study details}
\label{sec:studies}

Here we elaborate on the different technologies and respective studies contributing to the race. We start with a general introduction to each system with main references, links to website and source code if available, and design goals. Regarding implementation languages we distinguish between the user interface and low-level components. The hardware is described as either the conventional target processor or a custom architecture; for an overview refer to Fig.~\ref{fig:systems}. We also specify the parallelization model for software and hardware if applicable. A characterization of the employed spiking neural network simulation strategy includes the sequence of steps (e.g., code generation, compilation, network construction, and state propagation), but also details on algorithms and numerics. Each section concludes with details on power measurement.
These brief simulator profiles are by no means complete as some approaches are the results of decades of development and user experience. Features only touched upon may be the support for certain neuron and synapse models or plasticity mechanisms, or scalability with respect to network size. Our focus is on aspects that differentiate PD14 performance results on different systems, assuming that simulations achieve a sufficient accuracy. Where applicable, we highlight what we have learned from a computational point of view by contributing to the race.
Following the generic introductory paragraphs, we provide details on the particular PD14 results, i.e., the data points depicted in Fig.~\ref{fig:performance}, complementary to the performance data and system specifications in Table~\ref{tab:performance}.
For the complete study descriptions, we refer to the original publications.

With the maturation of the model specification, the studies employ two different sets of \nameref{sec:initialcond} resulting in statistically identical network activity but differing in the transient activity: the \textit{original initial conditions} used in~\citep{Potjans14_785} and \textit{amended initial conditions} suggested in~\citep{Rhodes19_20190160}.

\subsubsection{NEST CPU} \hfill\\
\label{sec:nest_cpu}

\noindent NEST~\cite{Gewaltig_07_11204} is a simulation software for spiking neural networks exploiting conventional CPU-based high-performance computing infrastructure. NEST is written in C++, uses OpenMP~\citep{OpenMPSpec} and MPI~\citep{MPIForum09} for parallelization, and has a Python user interface PyNEST~\citep{Eppler09_12,Zaytsev14_23} (website: \url{https://www.nest-simulator.org}, source code: \url{https://github.com/nest/nest-simulator}).
The code represents numerical values in double-precision floating-point format, and solves sub-threshold dynamics with exact integration~\cite{Rotter99a}. The network construction and simulation phase are natively executed in direct succession and both parallelized across all resources of the same system.
In the studies considered here, power of NEST simulations is measured with rack Power Distribution Units (PDUs), remotely reading the active power of the compute notes under load approximately once per second.

The race encouraged NEST developers to exploit the microscopic parallelism of the neuronal dynamics on many-core CPUs, including minimizing the number of barriers and employing thread-parallel memory allocation~\cite{Ippen2017_30}.

\begin{paragraph}{vAl+18~\cite{VanAlbada18_291}}
The study compares the accuracy and performance of NEST and SpiNNaker simulations by executing a custom PyNN~\cite{Davison09_10} code of the PD14 model with NEST~ 2.8~\cite{Nest280} using the original initial conditions. Simulations cover a duration of $T_\text{model}=10\,\text{s}$ and record spikes. The subsequent analysis considers the complete simulation phase and compares grid-based simulations with precise spike timing~\cite{Morrison07_47, Hanuschkin10_113} as a control of accuracy. In search for the optimal hardware configuration, the study carries out a strong-scaling experiment on an HPC cluster with 32 compute nodes (two CPUs per node). The system achieves the minimum $q_\mathrm{RTF}$ (\emph{vAl+18a}) on twelve compute nodes, and the minimum $E_\text{syn}$ (\emph{vAl+18b}) including an estimate for the network switch on three nodes.
\end{paragraph}
    
\begin{paragraph}{Kur+22~\cite{Kurth_2022}}
This work adapts the publicly available PyNEST~\cite{Eppler09_12} code of the PD14 model for benchmarking and uses the amended initial conditions. Runs of NEST~2.14.1~\cite{Nest2141} cover a duration of $T_\text{model}=10\,\text{s}$ following a warm-up time of $0.1\,\text{s}$, and do not record neuronal activity. A strong-scaling experiment scans a system of two HPC nodes with two CPUs per node hosting 64 cores each for optimal performance. The simulations reach the minimum $q_\mathrm{RTF}$ (\emph{Kur+22a}) on two point-to-point connected fully utilized nodes and the minimum $E_\text{syn}$ (\emph{Kur+22b}) already on one fully utilized compute node.
\end{paragraph}

\subsubsection{NEST GPU} \hfill\\
\label{sec:nest_gpu}

\noindent NEST GPU~\cite{Golosio21_627620, Tiddia22_883333} is an open-source simulator designed to simulate large-scale spiking neuron networks. Originally developed under the name NeuronGPU, it is now part of the NEST Initiative e.V. (source code: \url{https://github.com/nest/nest-gpu}). The simulator is written in CUDA-C++ and supports multi-GPU simulations through MPI. Moreover, it provides a Python interface, which closely mirrors that of the original NEST CPU code, allowing researchers to define neurons, connections, and synapse properties using similar commands.

The PD14 model size is too small to take advantage of the massive parallelism available at modern data centers, which offer researchers thousands of GPUs simultaneously. Yet, this model has been instrumental for investigating bottlenecks in single-GPU simulations, fostering the development of efficient algorithms for different simulation phases.

\begin{paragraph}{Gol+21~\cite{Golosio21_627620}}
This study presents the prototype library NeuronGPU, verifies the code in terms of simulation results, and validates performance by implementing the PD14 model and comparing with NEST CPU and GeNN using the amended initial conditions. The simulations span $T_{\text{model}}=10$\,s with an additional $1$\,s warm-up time at the beginning of each simulation to avoid startup transients. Spikes emitted after the first $1$\,s of activity are recorded for verification, whereas the recording was disabled to assess the library performance. Simulations are performed on Tesla V100 and GeForce RTX 2080Ti GPU cards, with the latter being the one with minimum $q_\mathrm{RTF}$ (\emph{Gol+21}). For the purpose of the present review, we revisit the configuration of \emph{Gol+21} and estimate $E_\text{syn}$ for the same workstation and software versions by measuring the wall socket power during the simulation phase. An earlier estimate of energy consumption for an analogous configuration, but only taking the GPU power consumption into account, comes from~\cite{kauth23_1144143}.
\end{paragraph}

\begin{paragraph}{Gol+23a~\cite{Golosio23_9598}}
The study presents a novel runtime network construction method for NEST GPU, enabling the execution of this phase entirely on the GPU hardware. In this framework, the PD14 is employed both for benchmarking and verification purposes. Regarding benchmarking, simulations are run for $T_{\text{model}}=10$\,s with an additional warm-up time of $0.5$\,s at the beginning of the simulation, amended initial conditions were used, and spike recording is disabled. The study performs the benchmarking on both data center and consumer GPUs (i.e., Tesla V100, Ampere A100, GeForce RTX 2080Ti, and GeForce RTX 4090). Among the different GPU hardware systems, the GeForce RTX 4090 achieves the minimum $q_\mathrm{RTF}$ (\emph{Gol+23a}). Here, we revisit the configuration of \emph{Gol+23a} and estimate $E_\text{syn}$ for the simulation codes and the workstation specified in \cite{Golosio23_9598}.
\end{paragraph}

\subsubsection{GeNN} \hfill\\
\label{sec:genn}

\noindent GPU enhanced Neuronal Networks (GeNN)~\cite{Yavuz16_18854} is a C++ library for generating efficient GPU kernels in CUDA or HIP for spiking neural network simulation (website: \url{https://genn-team.github.io}, source code: \url{https://github.com/genn-team/genn}). In more recent versions, this library is exposed to Python as PyGeNN~\cite{Knight21_15}.
GeNN is designed to be highly flexible, and one of the key enablers of this flexibility is that all neuron and synapse models are specified in a C-like language~(GeNNCode), which can be written directly in Python model descriptions.
While GeNN supports \emph{data-parallel} training of spiking machine learning models across multiple GPUs using NCCL~\cite{NCCL}, its current focus is performance on single GPUs. Spiking neural network simulations in GeNN comprise a code generation stage, compilation of generated C++ and CUDA/HIP code, followed by network initialisation and simulation. Network initialisation can largely be offloaded onto the GPU. Users of GeNN define neuron dynamics in terms of an update procedure for a timestep, meaning that the model definition already encompasses both the underlying differential equations and the numerical solver. 

\begin{paragraph}{KN18~\cite{Knight18_941}}
This study presents extensions to GeNN that support heterogeneous dendritic delays and on-GPU initialisation of state variables and connectivity (specified in GeNNCode). The authors benchmark a C++ implementation~\citep{knight_brainsonboardfrontiers_genn_paper_2018} of the PD14 model on GeNN 3.2.0~\citep{neworderofjamie_genn-teamgenn_2018} and a variety of GPU hardware (Jetson~TX2, GeForce~GTX~1050Ti, Tesla~K40c and Tesla~V100). The simulations use the original initial conditions, record all spikes, and the external drive consists of Poisson spike trains, directly delivered to each neuron. The benchmarks cover the duration of $T_\text{model}=10\,\text{s}$ and separately assess the wall clock time required  for initialization and state propagation. The authors record the power usage over time at the main socket and report energy to solution as well as energy per synaptic event. The fastest simulation is achieved on the V100 (\emph{KN18}). The investigation did not have physical access to the V100 system and therefore estimates power consumption by subtracting the GPU power reported by \texttt{nvidia-smi} on the K40-based system and adding the values for the V100.
\end{paragraph}

\begin{paragraph}{Kni+21~\cite{Knight21_15}}
This study presents a new Python frontend to GeNN~(PyGeNN) as well as a new GPU-side spike recording system.
In order to demonstrate the benefits of the new spike recording system and show that orchestrating simulations from Python has minimal performance overhead, the researchers benchmark GeNN 4.4.0~\citep{neworderofjamie_genn-teamgenn_2021} on some newer GPU hardware (GeForce~GTX~1650, Jetson Xavier~NX and  Titan~RTX) using a new PyGeNN implementation~\citep{knight_brainsonboardpygenn_paper_2021} of the PD14 model. The initialization strategy and external drive are the same as in \emph{KN18}. Simulations cover the duration of $T_\text{model}=1\,\text{s}$ and assess the wall clock time required for state propagation while recording all spikes. The fastest simulation is achieved on the Titan~RTX (\emph{Kni+21}).
\end{paragraph}

\begin{paragraph}{Gol+23b~\cite{Golosio23_9598}}
In the context of~\cite{Golosio23_9598}, the NEST GPU team considers GeNN as a reference and  performs analogous benchmarks with GeNN 4.8.0. Also for GeNN, the study achieves the best performance on the GeForce RTX 4090. Revisiting the earlier work we estimate $E_\text{syn}$ (\emph{Gol+23b}) using the same workstation as in~\cite{Golosio23_9598}.

\end{paragraph}

\subsubsection{SpiNNaker} \hfill\\
\label{sec:spinnaker}

\noindent SpiNNaker~\cite{furberpetrut} is a neuromorphic hardware platform designed to run spiking neural network simulations in biological real-time (source code: \url{https://github.com/spinnakermanchester}). This highly-parallel architecture uses low-power ARM-968 cores running at 200\,Mhz, with 64KB local data memory and 32\,KB local instruction memory, and 32-bit fixed-point calculations. Thus, there is no hardware support for floating-point operations. SpiNNaker chips made of up to 18 such cores are organized into boards of 48-chips which can be connected together. The biggest SpiNNaker machine built to date consists of over 1-million cores operating as a single machine. SpiNNaker machines execute spiking neural network simulations using the PyNN interface and sPyNNaker \cite{Rhodes18_816}, and SpiNNTools \cite{Rowley19_231} software APIs for facilitating problem description, partitioning, mapping, execution, and results extraction. To measure energy use in the studies considered here, a wall-socket power meter monitors an entire SpiNNaker system: SpiNNaker boards, communications switch, power supplies, and cooling fans. The meter used to profile PD14 simulations provides a reading accurate to 0.01\,kWh, with measurements taken at the beginning and end of simulations via software controlled readings. The authors subsequently convert this energy to solution into energy per synaptic event by dividing by the total number of synaptic events, and assuming constant power consumption throughout the simulation period~\cite{VanAlbada18_291, Rhodes19_20190160}.

Efforts to run the cortical microcircuit model have led to significant developments in the SpiNNaker software tool chain, and the PD14 model is now run regularly, in real-time and using the Poisson-drive, as part of the SpiNNaker software integration tests. This ensures that this model and others of a similar nature can be run on all future versions of the SpiNNaker software, demonstrating the lasting impact of the PD14 model on SpiNNaker development. The model has also helped inform the design of the successor system, SpiNNaker2~\cite{mayr2019spinnaker}, which is expected to deliver around a 10x improvement in functional density and energy-efficiency.

\begin{paragraph}{vAl+18c~\cite{VanAlbada18_291}}
This study reports the first successful execution of the PD14 model on SpiNNaker. The architecture was originally designed to run networks with 1\,ms timesteps in real-time, however the authors demonstrate that to accurately model the dynamics the PD14 model requires a simulation timestep of 0.1\,ms. While this validation was a key achievement, the resulting computation requires a slow-down of the machine to a real-time-factor of 20, due to the way the software environment maps the PD14 model to the hardware. The simulations use the original initial conditions as well as both DC (\emph{vAl+18c}) and Poisson drive, and gather energy and wall-clock time across: preprocessing, execution, and the extraction of results and subsequent post processing. The investigations highlight issues of the original software stack with the speed of loading data onto, and saving data from, SpiNNaker. This inspired development work enabling generation of synaptic data in parallel on the cores of the machine prior to execution, speeding up the loading process by several orders of magnitude, and significantly reducing the host memory requirements.
\end{paragraph}

\begin{paragraph}{Rho+19~\cite{Rhodes19_20190160}}
This study reports improvements to the SpiNNaker software stack to enable SpiNNaker to run the PD14 model in biological real time. It should also be noted this is a hard real-time solution, where every 0.1 ms simulation time-step was computed in 0.1 ms wall-clock time. This was achieved through revisions to the model itself, updating the initialization to reduce the number of neurons initialized with membrane potential state above threshold (leading to the amended \nameref{sec:initialcond}), and hence reducing excessive spike firing in the first timestep of the simulation, which had previously compromised the hard real-time compute requirements. Additional updates include revisions to the way computation is mapped to the SpiNNaker architecture, taking inspiration from previous research~\cite{Knight16_420, Galluppi15_429}, to enable parallelization of spike processing, a fundamental processing bottleneck in the study above~\cite{VanAlbada18_291}. Additional modifications improve synchronization of the SpiNNaker cores to ensure correct processing of the model, and include \emph{spike colouring} to identify delays of spikes being transmitted across the machine with the smaller timesteps in use. As  a result the study reports simulations with a real-time factor $q_{RTF}=1$ for both DC (\emph{Rho+19a}) and Poisson (\emph{Rho+19b}) drive, where energy measurements extend over 12 hours of wall-clock time -- an up-time with a consistent simulation never before achieved with SpiNNaker. 
\end{paragraph}

\subsubsection{CsNN} \hfill\\
\label{sec:csnn}

\noindent CsNN specifies a simulation architecture that can be prototypically mapped to an FPGA cluster as a hardware substrate. For CsNN, here the IBM INC-3000 system serves as the hardware platform. The IBM INC-3000 system consists of 16 PCB's (INC-cards, size 56\,cm  x 35\"cm), each hosting 27 reconfigurable Xilinx ZYNQ XC7Z045 SoC nodes. The whole INC-3000 system consists of 432 FPGA nodes which are connected by a 3-dimensional mesh. Earlier work already looked at single FPGAs (e.g., \cite{GuerreroRivera06_2651}). The IBM system, however, was initially part of the IBM General Artificial Intelligence project, in which networks of non-spiking neurons were investigated. It was then shown that this FPGA cluster architecture can also be used to advantage for the simulation of large-scale networks of spiking neurons.

\begin{paragraph}{Hei+22~\cite{heittmann22_728460}}
This study completely expresses the system architecture of CsNN in a high-level programming language and maps the specification to FPGA logic using high-level synthesis tools. This design flow enables the implementation and characterization of a large number of architecture variants as part of a broad design space exploration. The authors employ the framework to characterize different advanced ODE solvers in terms of their performance, with the perspective of integrating more realistic neuron and synapse models as used in PD14. Non-linear equations arise, for example, when considering bio-realistic conductance-based synaptic couplings, AdEx neurons~\cite{Brette-2005_3637}, Izhikevich neurons~\cite{Izhikevich03b}, or a non-linear membrane dynamics based on ion channels. In particular, all arithmetic operators are implemented in single-float precision. The work also extends the concept of procedural connectivity~\cite{Roth95_720,roth1997efficient} to the representation of synaptic connections with the attributes of synaptic weight and delay. Instead of storing synaptic parameters in a memory system, the algorithm derives all connection parameters at runtime by addressing individual seeds. The authors propose dedicated random number generators for the sampling from arbitrary distribution functions for the runtime-efficient online generation of synaptic parameters. The approach maps the connectivity of PD14 to a highly parallel near-memory computing architecture with a high compression factor and rapid look-up of individual connections.

In the simulation, the authors parameterize online generators for the network generation in such a way that they exactly reproduce the detailed connection statistics of the genuine PD14 model. The algorithm not only reproduces network densities but also the properties of synaptic multapses (multiple connections between pairs of neurons \cite{Senk22_e1010086}). The study avoids initial transients by determining  the initial conditions for the membrane potentials, synapse states, and the delay buffers from the steady-state  of a prior test simulation. Manufacturer data sheets provide first worst-case estimates of energy consumption for CsNN~\cite{kauth23_1144143}. For the present review, we revisit the original configuration and determine energy consumption in simulations covering an interval of $T_\text{model}=15\,\text{min}$ (\emph{Hei+22}).
\end{paragraph}

\subsubsection{neuroAIx} \hfill\\
\label{sec:neuroaix}

\noindent neuroAIx is a simulation platform for computational neuroscience, comprising a cluster of 35 FPGAs~\cite{kauth23_1144143}. Its development was driven by identifying key requirements of such a system: flexibility, observability, scalability and replicability. The study first analyzes the computational properties of biological networks of spiking neurons in the human brain. This results in an application-specific architecture for tackling the three key bottlenecks limiting the performance of previous platforms: computation, memory and communication~\cite{kauth23_1144143}. Firstly, to solve the computational bottleneck, neuroAIx implements two separate solutions for computing neuronal state updates at extremely low latencies: a dedicated register-transfer level design for prevalent neuron models like leaky integrate-and-fire, and the custom neuron processor architecture nAIXt~\cite{Kauth24_3}. Secondly, the architecture tackles the memory bottleneck by directly accessing off-chip memory bypassing any processor cores and with latency hiding techniques like prefetching synapse data. Thereby, the system avoids latency overheads (beyond the technical limitations of the underlying memory modules). Thirdly, regarding the communication bottleneck, the authors introduce overlayed long hop topologies~\cite{Kauth20_1} as an extension to mesh-like network topologies. Their key features are symmetric connections between distant network nodes to minimize the maximum number of hops needed to communicate spike messages. The study combines long hops with a customized routing algorithm and deadlock prevention~\cite{Sobhani22_70}, a condensed version of the Go-Back-N ARQ flow control protocol and a synchronization scheme ensuring minimum latency.

The resulting platform can replay simulations and produce bit-equivalent results, while exhibiting spiking behavior statistically equivalent to simulations in NEST. First analyses additionally confirm weak and strong scaling properties. Future iterations of neuroAIx will further increase its performance and applicability, by building on newer FPGAs, and introducing support for (three-factor) plasticity rules and a user-friendly cloud interface with NEST/NESTML integration.

\begin{paragraph}{Kau+23~\cite{kauth23_1144143}}
The work introduces the neuroAIx FPGA cluster and demonstrates its performance on relevant neural network sizes using the PyNEST PD14 model \cite{Eppler09_12} with NEST v3.3~\cite{Nest33} as a reference. This code includes the original initial conditions and uses DC input to drive activity. For comparability, the study uses the same generated connectome for simulations on neuroAIx and an HPC cluster. Finally, a simulation covering  a duration of $T_\text{model}=15\,\text{min}$ assesses performance, where it discards the first second to ignore transient activity. All spikes are recorded and transmitted to a host computer to verify the resulting dynamics using second-order statistics like inter-spike intervals (with spike recording on neuroAIx incurring no performance hits). neuroAIx achieves the minimum $q_\mathrm{RTF}$ by using the entire cluster of 35 FPGAs. The authors estimate $E_{\text{syn}}$ by measuring the power consumption of individual FPGAs during simulation using current clamps at the wall socket and verify results against data reported by the internal power management units.
\end{paragraph}

\subsubsection{BrainScaleS} \hfill\\
\label{sec:brainscales}

\noindent The physical implementation of neuron and synapse dynamics in analog microelectronic circuits can offer significant advantages in energy efficiency and speed compared to numerical simulation. BrainScaleS-1~\cite{Schmidt23_13} is a mixed-signal wafer-scale accelerator (implemented in \SI{180}{\nano\meter} CMOS) for plastic spiking neural networks providing high-speed model dynamics and energy-efficient operation. A multi-layered software stack~\cite{Mueller22_790} provides automated conversion between user-defined experiments in the high-level description language PyNN and the neuromorphic substrate.

\begin{paragraph}{Sch+25~\cite{Schmidt25_nice}}
The study implements the PD14 model on a single BrainScaleS-1 wafer-scale module, featuring 196k AdEx neuron circuits and 43M plastic conductance-based synapses. The authors adapt the model to fit the hardware's properties, including down-scaling to accommodate the biological fan-in following the approach outlined in~\cite{Albada15}:
In particular, the number of synapses in the PD14 model is greater than the number of available synapses on the silicon substrate.
In addition, biological fan-in, or connection density, requires BrainScaleS neuron circuits to be combined, thereby decreasing the number of available model neurons per wafer module.
The modified model represents the external drive by DC current.
Additionally, the work replaces the original current-based synapse model to match the substrate's conductance-based dynamics.
This adaptation process, transitioning from a numerical model to a neuromorphic implementation, is non-trivial, but promises order-of-magnitude improvements in energy efficiency and latency compared to conventional software simulation as well as hardware-accelerated numerical simulation.
For the down-scaled model implementation running at a real-time factor of $q_\text{RTF} = 0.0001$ the system achieves a peak rate of $162 \times 10^9$ spikes per second and an upper bound of $E_\text{syn} < \SI{0.012}{\micro\joule}$.
\end{paragraph}

\subsection{Benchmarking recipe}
\label{sec:recipe}

The joined experience of the individual \nameref{sec:studies} reviewed here suggests guidelines for the evaluation of the performance of computing systems using the PD14 model as a benchmark.
The steps to be taken are: 1) to obtain the reference model from a reliable source, 2) to implement a simulator-specific algorithmic interpretation of the model, 3) to achieve sufficient accuracy, and finally 4) to assess performance in terms of time and energy to solution (compare with Fig.~\ref{fig:graphical_abstract}). Here we elaborate on the reasoning behind simulation and parameter recommendations and provide a checklist in Table~\ref{tab:recommendations}.

\begin{table}[h]
\centering
\begingroup
\definecolor{darkcyan}{HTML}{225555}
{\rowcolors{1}{darkcyan!10}{darkcyan!20}
\renewcommand{\arraystretch}{2}
{
\begin{tabular}{p{4cm}|p{10cm}}
\toprule
Model reference & \url{https://github.com/INM-6/microcircuit-PD14-model}\\
\hline
External drive & State whether DC or Poisson is used\\
\hline
Initial conditions & Amended initial conditions: distribute membrane potentials normally with population-specific mean and variance\\
\hline
Warm-up time  & Discard the initial 500\,ms of model time from the data to be analyzed\\
\hline
Simulation duration & Accuracy: $T_\text{model}=15\,\text{min}$, performance:  $T_\text{model}\ge10\,\text{s}$\\
\hline
Repeated simulations & Average across ten random seeds\\
\hline 
Spike recording & Accuracy: yes, performance: no\\
\hline
Accuracy & Compute distributions of 1) single-neuron \textit{firing rate} (FR), 2) \textit{coefficient of variation} (CV) of the inter-spike intervals (ISI), and 3) short-term spike-count \textit{correlation coefficients} (CC), and compare with reference data\\
\hline
Performance & Measure real-time factor $q_\text{RTF}$ and the energy per synaptic event $E_\text{syn}$ (include all contributions necessary for running the simulations at the power outlet)\\
\bottomrule
\end{tabular}
}}
\endgroup
\caption{Checklist with recommended model and simulation parameters for the PD14 model.}
\label{tab:recommendations}
\end{table}

\subsubsection{Model reference} \hfill\\
\label{sec:recipe_reference}

The original neuroscientific publication introducing the PD14 model and describing the data integration process as well as the neuroscientific conclusions is \cite{Potjans14_785}. However, on the basis of the progress of the community in the formal description of neuronal network models since the first publication of PD14  (including the scientific case for large-scale brain simulations~\cite{Einevoll19_735}, connectivity concepts~\cite{Senk22_e1010086}, and the modeling language NESTML~\cite{Linssen25_1544143}) and the results of the present study, we recommend to use the novel model reference:
The open access web site \url{https://github.com/INM-6/microcircuit-PD14-model} provides a detailed, implementation-agnostic mathematical description of the model, as well as a documented PyNEST reference implementation. The repository also collects the performance results of this study and supports the inclusion of new data points as they become available.

\newpage

\subsubsection{Model and simulation parameters} \hfill\\
\label{sec:recipe_parameters}

\begin{paragraph}{External drive}
In the PD14 model, each neuron receives input spikes from an independent Poisson process with a population-specific rate.
This rate is parameterized by a global base rate multiplied by a population-specific in-degree. According to the original study~\cite{Potjans14_785}, the Poisson input can be replaced by a DC input corresponding to the mean current delivered by the Poisson input (see their Figure 7) without compromising the accuracy of the network activity statistics. While Poisson input might be biologically more realistic, DC input is less computationally expensive and avoids the dependence of the performance data on the efficiency of the implementation of  the random number generators. Some studies considered in this review use Poisson and others DC input as stated in Table~\ref{tab:performance}. We recommend to clearly state which external drive is used for performance comparability.
\end{paragraph}

\begin{paragraph}{Initial conditions}
\label{sec:initialcond}
Simulating neurons firing at higher rates (also temporarily) requires more communication and is, for event-driven algorithms, therefore computationally more expensive than at lower rates. The dynamics of the PD14 model are mostly stable with only weak fluctuations in global and per-neuron firing rates~\cite{Dasbach21_90}. Convergence speed into this state is determined by initial conditions. The original study~\cite{Potjans14_785} uses normally distributed initial membrane potentials with the same mean and standard deviation for all populations. Here we recommend to distribute initial conditions with the population-specific means and variances of the stationary network state after initial transients have decayed~\cite{Rowley15_19230}. The present work calls this set of parameters the amended initial conditions. The stable network states resulting from the two sets are statistically identical. While avoiding transient activity with high firing rates, population-specific distributions of initial membrane potentials also ensure fast convergence to the stable state.
\end{paragraph}

\begin{paragraph}{Warm-up time}
To avoid analyzing data from initial transients in the dynamics of the model, we recommend discarding a warm-up time of $500\,\text{ms}$. This holds for both the statistical analysis of spike data and the measurement of simulation speed.
\end{paragraph}

\begin{paragraph}{Simulation duration}
The specificity of verification measures relies on the observation duration. For an investigation of this relationship on the example of the PD14 model see~\cite{Dasbach21_90}. Simulation times need to be large enough to distinguish model specific spike correlations from spurious correlations of independent spike trains of finite length. To assess accuracy, we recommend simulating for a biological time of $T_\text{model}=15\,\text{min}$. To assess performance, we recommend simulating for at least $T_\text{model}=10\,\text{s}$. An increased speed of the simulation  requires a longer stretch of biological time covered by the model to keep the time it takes the computing system to complete the simulations large enough for reliable measurements of wall-clock time and energy consumption.
\end{paragraph}                      
         
\begin{paragraph}{Repeated simulations}
The PD14 model is a highly irregular system and the network activity is evaluated on a statistical level. Random numbers are involved in drawing connections during network construction, initializing membrane potentials, and possibly providing Poisson input during the simulation. We recommend averaging simulation results across at least ten random seeds to gain statistics across a range of microscopically different network instantiations and dynamics. In addition, we also recommend repeating simulations for performance measurements to account for fluctuations in the computing system.
\end{paragraph}
     
\begin{paragraph}{Spike recording}
Recording spike data to files or transfer of the data to a companion job is required for accuracy assessment. A  simulation engine may buffer spikes locally and only write them to files after a data limit has been reached. Consequently, the performance of the simulation may depend on the stretch of time covered by the simulation. To disentangle the performance of the simulation engine from the performance of analysis backends, we suggest not to record spikes when assessing the performance.
\end{paragraph}

\subsubsection{Accuracy} \hfill\\
\label{sec:recipe_accuracy}

\noindent We recommend a statistical comparison of the obtained spike data to reference data created by a trusted simulation code.
The distributions of single-neuron \textit{firing rate} (FR), distributions of \textit{coefficient of variation} (CV) of the inter-spike intervals (ISI), and distributions of short-term spike-count \textit{correlation coefficients} (CC) should be computed. For details, refer to~\cite[Sec. 2.3.1]{Dasbach21_90} and see also~\cite{Gutzen18_90}. 
Demonstrating the visual overlap of these distributions with the reference is the minimum requirement. Computing scores to quantify the match of the distributions is appreciated such as Kullback-Leibler divergence, Kolmogorov-Smirnov test, or the Earth Mover Distance.

\subsubsection{Performance} \hfill\\
\label{sec:recipe_performance}

\noindent Time to solution and energy to solution should be assessed by the real-time factor and the energy per synaptic event, respectively. For definitions, see the beginning of \nameref{sec:results}. In general, we recommend that power measurements account for the total system power: all contributions necessary for running the simulations measurable at the power outlet. This definition excludes additional potential energy costs stemming, for example, from keeping the computing system at a stable temperature. While important for assessing the true cost for the operator of a computing facility, it is a not required to assess the power of the different systems in a pragmatic and comparable fashion.

\funding{
This project received funding from
the European Union’s Horizon Europe Programme under the Specific Grant Agreement No. 101147319 (EBRAINS 2.0 Project);
the Joint Lab “Supercomputing and Modeling for the Human Brain”;
Juelich Research Centre intramural STEF fund for the update of instruments;
HiRSE\_PS, the Helmholtz Platform for Research Software Engineering - Preparatory Study, an innovation pool project of the Helmholtz Association;
the Helmholtz Association’s Initiative and Networking Fund under project number SO-092 (Advanced Computing Architectures, ACA);
the Federal Ministry of Education and Research (BMBF, Germany) through the projects NEUROTEC II (grant number 16ME0399) and Clusters4Future-NeuroSys (grant number 03ZU1106CA);
EPSRC (grant numbers EP/P006094/1, EP/S030964/1 and EP/V052241/1);
the European Union’s Horizon 2020 research and innovation program under Grant Agreement 945539 (HBP SGA3); and Google Summer of Code.
The contribution of the University of Cagliari to this study was supported by the Project e.INS Ecosystem of Innovation for Next Generation Sardinia—spoke 10-CUP F53C22000430001—MUR Code No. ECS00000038.
Authors involved in NEST GPU development acknowledge the use of Fenix Infrastructure resources, which are partially funded from the European Union’s Horizon 2020 research and innovation programme through the ICEI project under the Grant Agreement No. 800858.
}

\printbibliography

\end{document}